\documentclass[12pt,preprint]{aastex}

\begin{document}

\title{Early Fermi Gamma-ray Space Telescope Observations \\
  of the Quasar 3C~454.3}


\author{
A.~A.~Abdo\altaffilmark{1,2},
M.~Ackermann\altaffilmark{3},
M.~Ajello\altaffilmark{3},
W.~B.~Atwood\altaffilmark{4},
M.~Axelsson\altaffilmark{5,6},
L.~Baldini\altaffilmark{7},
J.~Ballet\altaffilmark{8},
G.~Barbiellini\altaffilmark{9,10},
D.~Bastieri\altaffilmark{11,12},
M.~Battelino\altaffilmark{5,13}
% ...
}
\altaffiltext{1}{National Research Council Research Associate}
\altaffiltext{2}{Space Science Division, Naval Research Laboratory, Washington, DC 20375}
\altaffiltext{3}{W. W. Hansen Experimental Physics Laboratory, Kavli Institute for
Particle Astrophysics and Cosmology, Department of Physics and SLAC National Accelerator
Laboratory, Stanford University, Stanford, CA 94305}
\altaffiltext{4}{Santa Cruz Institute for Particle Physics, Department of Physics and
Department of Astronomy and Astrophysics, University of California at Santa Cruz, Santa
Cruz, CA 95064}
\altaffiltext{5}{The Oskar Klein Centre for Cosmo Particle Physics, AlbaNova, SE-106 91
Stockholm, Sweden}
\altaffiltext{6}{Department of Astronomy, Stockholm University, SE-106 91 Stockholm,
Sweden}
\altaffiltext{7}{Istituto Nazionale di Fisica Nucleare, Sezione di Pisa, I-56127 Pisa,
Italy}
\altaffiltext{8}{Laboratoire AIM, CEA-IRFU/CNRS/Universit\'e Paris Diderot, Service
d'Astrophysique, CEA Saclay, 91191 Gif sur Yvette, France}
\altaffiltext{9}{Istituto Nazionale di Fisica Nucleare, Sezione di Trieste, I-34127
Trieste, Italy}
\altaffiltext{10}{Dipartimento di Fisica, Universit\`a di Trieste, I-34127 Trieste, Italy}
\altaffiltext{11}{Istituto Nazionale di Fisica Nucleare, Sezione di Padova, I-35131
Padova, Italy}
\altaffiltext{12}{Dipartimento di Fisica ``G. Galilei", Universit\`a di Padova, I-35131
Padova, Italy}
\altaffiltext{13}{Department of Physics, Royal Institute of Technology (KTH), AlbaNova,
SE-106 91 Stockholm, Sweden}
% ...


\begin{abstract}
This 
is the first report of Fermi Gamma-ray 
Space Telescope observations of the quasar 
3C~454.3, which has been undergoing 
pronounced long-term outbursts since 2000. 
The data from the Large Area Telescope (LAT),
covering 2008 July 7--October 6, indicate strong, highly
variable $\gamma$-ray emission with an average flux 
of $\sim 3 \times 10^{-6}$ photons cm$^{-2}$ s$^{-1}$, for energies 
$>100$\,MeV. The
$\gamma$-ray flux is variable, with strong, distinct,
symmetrically-shaped flares for which the flux increases by a factor of
several on a time scale of about three days.  This variability indicates
a compact emission region, and the requirement that the source is
optically thin to pair-production implies relativistic beaming with
Doppler factor $\delta > 8$, consistent with the values
inferred from VLBI observations of superluminal expansion ($\delta \sim 25$).  
The observed $\gamma$-ray
spectrum is not consistent with a simple power-law, but instead steepens
strongly above $\sim 2$ GeV, and is well described by a broken power-law
with photon indices of $\sim 2.3$ and $\sim 3.5$ below
and above the break, respectively.  This is the first direct observation of 
a break in the spectrum of a high luminosity blazar 
above $100$\,MeV, and it is likely 
direct evidence for an intrinsic break in the energy distribution of the 
radiating particles. Alternatively, the spectral softening above 2\,GeV 
could be due to $\gamma$-ray absorption via  photon-photon pair production
on the soft X-ray photon field of the host AGN, but such an interpretation 
would require the dissipation region to be located very close ($\lesssim 100$
gravitational radii) to the black hole, which would be inconsistent with 
the X-ray spectrum of the source.  
\end{abstract}


\section{Introduction}

The successful launch on 2008 June 11 of the Gamma-ray Large Area Space
Telescope (GLAST) --- now known as the Fermi Gamma-ray Space
Telescope --- ushered in a new era of observational astronomy in the
energetic $\gamma$-ray band.  Early observations (including those
conducted during in-orbit checkout) proved that the performance
of the Large Area Telescope (LAT; 
for details see Atwood et al. 2009), sensitive in
the band $20$ MeV to $> 300$ GeV, has been close to the
pre-flight predictions (Abdo et al. 2008).  The data obtained with the
EGRET instrument on-board the Compton Gamma-ray Observatory, LAT's
predecessor, indicated that the most prominent extragalactic energetic
$\gamma$-ray sources are blazars, a sub-class of active galactic
nuclei whose overall flux is dominated by emission from a 
relativistically boosted inner ($\leq$\,pc) jet. Here, we use an 
example of early LAT observations of the blazar, 3C~454.3, to highlight 
the capabilities of the instrument, but also to derive further constraints 
on the emission mechanisms and structure of the object.

Observationally, blazars are characterized by large
amplitude chaotic variability measured in all accessible spectral
bands, from radio to GeV or even TeV energies. 
The variability often manifests itself as very high flux states
that last for months to years, with more rapid, smaller amplitude 
flares superimposed on those high states.  
Optical and radio data
show a high degree of polarization, and the radio data reveal
the presence of strong emission components that arise from extremely compact
($\sim$ milliarcsec), spatially and spectrally variable structures 
with a core-jet morphology, often associated with apparent superluminal 
expansion.  

3C~454.3, at  redshift $z=0.859$, is a well-known example of this class 
of objects.  Long-term VLBI monitoring indicates a jet Lorentz factor 
of $\Gamma_{\rm jet} = 15.6 \pm 2.2$, an angle of motion to 
our line-of-sight of $\theta = 1.3 \pm 1.2^\circ$, and a corresponding
Doppler factor of $\delta \equiv (\Gamma_{\rm jet} (1-\beta \, \cos
\theta))^{-1} \sim 25$ (Jorstad et al. 2005).
These values are consistent with recent work by Lister et al. (2009). 
This object entered a high-flux phase in 2000 and was remarkably 
active in 2005, when it reached the largest apparent optical 
luminosity ever recorded from an astrophysical source apart from GRBs 
(Fuhrmann et al.  2006; Villata et al. 2006). The 2005 outburst was covered 
also in the X-ray range by the Swift satellite (Giommi et al. 
2006). Unfortunately, due to the lack of contemporaneous $\gamma$-ray data, no 
firm conclusions could be drawn regarding the bolometric energetics of 3C~454.3 
during this outburst. Historically, the source was detected above 100 MeV 
by EGRET (Hartman et al. 1999), and in the softer $\gamma$-ray bands by OSSE 
(McNaron-Brown et al. 1995) and COMPTEL (Zhang et al. 2005). The $\gamma$-ray 
flux measured by the AGILE satellite in 2007, when the source was optically 
fainter than in 2005 but much brighter than during its low states (Ghisellini 
et al. 2007; Vercellone et al. 2008; 2009), was much higher than the 
$\gamma$-ray  flux recorded by EGRET during low states (see Raiteri 
et al. 2008 for multi-band data). This indicates that the synchrotron 
(optical) outbursts in 3C~454.3 are indeed generally accompanied by 
$\gamma$-ray outbursts. 
In particular, the AGILE observations found an average flux of 
$F_{\rm E>100 MeV} = 2.8 \pm 0.4 \times 10^{-6}$\,photons\,cm$^{-2}$\,s$^{-1}$ 
in 2007 July 24--30 (Vercellone et al. 2008), and a flux of
$1.7 \pm 0.13 \times 10^{-6}$\,photons\,cm$^{-2}$\,s$^{-1}$ 
in 2007 November 10--December 1, 
with the spectrum well-described by a power 
law with photon index $\Gamma = 1.73 \pm 0.16$ (Vercellone et al. 2009).  
The MAGIC telescope also observed but did not detect 3C~454.3 at TeV 
energies during these epochs;  the non-detection implies 
a spectral cut-off at $\sim 20$ GeV (Anderhub et al. 2008).  

As expected, 3C~454.3 was detected easily by Fermi (Tosti et
al. 2008).  Owing to its high flux state, it was possible for the
LAT to measure its variability properties on time scales of less than a
day. The $\gamma$-ray observations by the LAT are described in
Section 2. Given the strong and variable $\gamma$-ray flux,
simultaneous data were highly desirable, and we secured a
number of observations in the radio, optical, and X-ray bands.  We
report on some of those results here, although more detailed inter-band
timing correlations will be the subject of a future paper. In Section 3,
we present the overall spectrum of the source and compare the
$\gamma$-ray variability properties to those measured in other bands.
There, we also present an emission model for the source and 
highlight the new constraints derived from our data. We summarize 
our findings in Section 4.  

\section{Observations}

\subsection{LAT Data: Light curve and $\gamma$-ray spectrum}
The LAT (Atwood et al. 2009) simultaneously monitors $\sim
2.4$ steradians on the sky, albeit with an effective area that varies
significantly with the arrival direction of the incident photon with
respect to the instrument pointing direction.  In survey mode, the instrument z-axis
is offset $35^\circ$ North and South from the spacecraft zenith during
alternate orbits in order to provide complete sky coverage every three
hours.  The LAT instrument was turned on 2008 June 25, and the data
presented here were collected during the early check-out phase (which
ended on August 11) until October 6.  Most data were taken in survey
mode, the source being observed at an average off-axis angle of $37^{\rm o}$. 
The effective area decreases to about 80\% and 70\% of the on-axis value at  
$30^{\rm o}$ and  $40^{\rm o}$ respectively. About 7000 photons with E $>$ 100 MeV 
ascribed to 3C~454.3 were detected during the continuous survey mode extending from 
August 4 to October 6.  The exposure at 1 GeV for this period was
$\simeq$ 6.5$\times 10^5$ m$^2$s.

For the period July 7--July 31, the instrument was
alternately pointed at two sky regions centered on Vela and
3EG~J1835+5918. In the later case, 3C~454.3 was observed with an
average off-axis angle of $55^{\rm o}$, i.e., close to the edge of the
field-of-view, with a corresponding reduction of the effective area
of about 50\% with respect to the on-axis value. 

The light curve obtained with the LAT is shown in Figure 1.  
We used the standard LAT analysis software, {\tt
ScienceTools v9r7}\footnote{http://fermi.gsfc.nasa.gov/ssc/data/analysis/documentation/Cicerone/}, 
and performed a maximum likelihood fit of the model parameters.  The source 
model included a point source for 3C~454.3, a component for the Galactic 
diffuse emission derived using the GALPROP code (Strong, Moskalenko, \& Reimer 2004; Strong et
al. 2004), and an isotropic component to represent, in combination,
the extragalactic diffuse emission and the residual instrument
background.  The spectral shape of the isotropic component was derived
from high Galactic latitude ($|b| > 60^\circ$) sky survey data
accumulated over a similar time period as the data for 3C~454.3.  The
events were selected using the most stringent set of standard analysis
cuts and correspond to the so-called ``diffuse class'' events, which
comprise the highest quality photon events in terms of direction and
energy reconstruction and background rejection.  The events were
extracted in the range 100 MeV--300 GeV and within a $15^\circ$
acceptance cone centered on the location of 3C~454.3.  This region is
substantially larger than the 68\% containment angle of the PSF at the
lowest energies ($\sim 3^\circ$). This is necessary in order to
constrain the diffuse emission components accurately.  Since the
likelihood calculation models the spatial distribution of events as a
function of energy, this procedure naturally accounts for the narrower
PSF at higher energies when fitting the point source spectral
parameters.  
  
We note that 3C~454.3 was sufficiently bright that the
background contribution within a few degrees of the source is only a
small fraction of the source count rate and that the observed 
flux variations of the source do not correlate with the changing background 
level along the satellite orbit.  Our current estimate of 
systematics is $<20\%$ on the flux.    

Figure 1 shows that the source is consistently
variable, with the rise and decay times of $\sim 3.5$ days,
corresponding to the doubling of the flux in 2 days.  
The spectral shape appears constant during this period 
within the systematic uncertainties. 
The flux $F_{\rm E>100 MeV}$ peaks at roughly $1.2 \times 10^{-5}$ 
photons cm$^{-2}$ s$^{-1}$ on MJD 54657 (2008 July 10). 
Note that such an exceptional flux from any blazar 
was only observed twice during the EGRET era, for
3C~279 (Wehrle et al. 1998) and PKS~1622-297 (Mattox et al. 1997).
In survey mode, the relative change of reconstructed flux due to
uncorrected acceptance variations (calibration of the effective area
as a function of the off-axis angle) has been determined to be about
5\% for a daily light curve using the data from the (steady) bright pulsars.
An additional 15\% uncertainty has been estimated to affect the data
collected in pointing mode presented here, because of the proximity
of the source direction to the edge of the field of view. 

We have also fit the broad-band spectrum of the combined data from the
onset of the regular survey mode, August 3, through September 2. As
with the daily light curve analysis, we have used the standard
``diffuse'' class event selections and restricted the energy range to
200~MeV--300~GeV.  Since 3C~454.3 has consistently been one of the
brightest sources since the beginning of the mission, the data
accumulated over the month of survey data are sufficient to allow us
to consider models that are more complex than a simple power-law in
trying to fit the source spectrum.
 
Our basic model does use a simple power-law for 3C~454.3, and that model 
yields a photon index 
of $\Gamma = 2.40 \pm 0.03_{\rm stat} \pm 0.09_{\rm sys}$. 
 This is substantially {\em softer} than the index reported by AGILE,  but the
data were not contemporaneous. The difference could 
also be partially due to different bandpasses of the two
instruments.  The residuals to the power-law fit clearly indicate that
the spectral model for the $\gamma$-ray emission from 3C~454.3 must be
more complex; and in particular, the data show a steepening towards
higher energies.  A broken power-law model yields $\Gamma_{\rm low} =
2.27 \pm 0.03_{\rm stat} \pm 0.09_{\rm sys}$, $\Gamma_{\rm high} = 3.5
\pm 0.2_{\rm stat} \pm 0.15_{\rm sys}$, and a break energy of $2.4 \pm
0.3_{\rm stat} \pm 0.3_{\rm sys}$ GeV.  The likelihood ratio test 
gives the probability for
incorrectly rejecting the power-law model in favor of the broken
power-law model as $5 \times 10^{-12}$. The break was consistently detected 
at similar energies when repeating the analysis for different one-month 
periods (Aug., Sep., Oct.). An instrumental effect can safely be ruled 
out, given the level of the systematic uncertainties.  An unfolded $\nu F_\nu$
spectrum corresponding to this model is shown in Figure 2.  We note
that with the current data, we cannot distinguish between various
models describing the steepening spectrum, e.g., broken power-law
vs. exponentially cut-off power-law, but such differentiation might be
possible with more data.  Nonetheless, this is the first indication 
of an observed break in the $\gamma$-ray spectrum of a high luminosity
blazar, calling for a complex spectral model beyond a simple power-law
approximation.  With this model, we determine the time-averaged flux 
for the 2008 August 3 -- September 4 epoch to be 
$F_{\rm E>100MeV} = 3.0 \pm 0.1 \times 10^{-6}$ photons cm$^{-2}$ s$^{-1}$.

\subsection{Low Energy Observations}

Triggered by the detection of a high flux level during the first
intensive Fermi-LAT observations of 3C~454.3 in 2008 July/August, the
broad-band behavior of the source was monitored via an ad-hoc
multi-wavelength (MW) campaign covering the period 2008 July to 2009 January. 
These intensive observations involved ground-based monitoring at
radio cm/mm/sub-mm wavelengths (e.g., Effelsberg 100-m, IRAM 30-m, SMA and 
OVRO 40-m telescopes) and at IR/optical bands (e.g., Spitzer, 
Kanata/Hiroshima 1.5-m and REM telescopes, plus 
telescopes of the GASP-WEBT consortium).
In addition, daily 2 ksec observations were performed with the Swift
satellite, providing X-ray data, as well as optical and
UV coverage.  Details of this multi-wavelength campaign, 
including all collected data and
the corresponding MW analysis of the source behavior between 2008 June
and 2009 January at radio/IR/optical/UV/X-ray/$\gamma$-ray bands, will
be presented in a subsequent paper (Abdo et al. 2009, in preparation). 
For the purpose of the first LAT results presented here, 
we will use a sub-set of these MW data to
construct a quasi-simultaneous spectral energy distribution of
3C~454.3 including radio, optical, UV, X-ray and LAT $\gamma$-ray data. 
In Figure 3, we show the corresponding data collected during the 
period MJD 54685--54690 (2008 August 7--12). 

The radio cm/mm band data were obtained with the Effelsberg 100-m and 
IRAM 30-m telescopes in the framework of a Fermi-related monitoring
program of potential $\gamma$-ray blazars (F-GAMMA project, Fuhrmann et al. 2007). 
At Effelsberg, observations were conducted quasi-simultaneously at eight frequencies 
between 2.6 and 42\,GHz using cross-scans in azimuth and elevation direction. 
Consequently, pointing off-set correction, gain correction, atmospheric opacity 
correction and sensitivity correction have been applied to the data (for details 
see Fuhrmann et al. 2008 and Angelakis et al. 2008). The observations at the IRAM 30-m 
telescope were obtained with SIS receivers (operating at 86, 142 and 228\,GHz) and 
with calibrated cross-scans in azimuth and elevation direction. The receiver calibration
was done using hot and cold loads (standard chopper wheel method) and the opacity corrected 
intensities were converted into the standard temperature scale. After Gauss fitting of 
the averaged sub-scans, each temperature was corrected for remaining pointing offsets 
and systematic gain-elevation effects. The conversion to the standard flux density 
scale was done using the instantaneous conversion factors derived from 
primary calibrator measurements (for details, see Ungerechts et al. 1998).  

The optical, UV, and soft X-ray data were obtained from 2 ksec pointings with the Swift 
satellite as a part of daily TOO observations during the campaign. For the 
screening, reduction and analysis of the data from Swift instruments we used standard 
procedures within the HEASoft v. 6.5 software package with the calibration 
database updated as of 2008 June 26. The XRT operated in photon 
counting mode, and the analysis was performed with the {\tt xrtpipeline} 
task with default parameters and having selected photons with grades 0--12. 
We also checked for the presence of pile-up. The UVOT data were integrated 
with the {\tt uvotimsum} task and analyzed with the {\tt uvotsource} 
task using source region radii of $5''$ for the optical filters and $10''$ 
for the UV filters, respectively. The background was extracted from a nearby source-free 
circular region with $50''$ radius. The observed magnitudes were converted 
into flux densities according to standard procedures (Poole et al. 2008).

\section{Discussion}

\subsection{Overall Spectral Energy Distribution and Modeling of the Source}

The amplitude of blazar variability can be as high as a factor of 100,
so many important inferences with regard to the source structure 
can be derived most robustly from broad band
spectra obtained simultaneously, with additional, crucial constraints
from time-resolved broad-band spectroscopy.  To this end, we assembled
the data as given above, contemporaneous with our 2008 August epoch. 
We plot a quasi-simultaneous spectral energy distribution (SED)
using our MW data from August 7--12 in Figure 3.  The overall SED
of the source is broadly similar to that observed
for other EGRET blazars, revealing two broad peaks: one in the mm-to-IR
band and the other in the $\gamma$-ray band.  Given the polarization
measured in all accessible segments of the low energy component, the
most viable emission mechanism is the synchrotron process, produced by a
distribution of relativistic electrons radiating in an ambient
magnetic field.  

The high energy peak, in turn, could be due to Compton scattering of 
either the synchrotron photons internal to the jet (see, e.g., 
Ghisellini, Maraschi, \& Treves 1985) or the radiation 
associated with the nucleus, such as the photons produced by the 
accretion disk, broad emission line region, or the thermal emission produced 
by the circumnuclear dust in the host galaxy (Dermer, Schlickeiser, 
\& Mastichiadis 1992;  Sikora, Begelman, \& Rees 1994).
Radiation from ultra-high energy protons or ions that interact with
magnetic fields associated with the jet or its environment, including
subsequent cascades, may be an alternative explanation for the high
energy component (e.g., M\"ucke et al. 2003; Atoyan \& Dermer 2003;  
B\"ottcher, Reimer, \& Marscher 2008).  

Inferences regarding the source structure are
strongly model-dependent, even within leptonic emission models. A
comparison of results presented for 3C~454.3 in Vercellone et al. (2009)
with the analysis of the broad-band temporal behavior of Sikora et al. (2008)
clearly indicates that a wide range of parameters can describe 
the same data. Nonetheless, the 
different models do find roughly consistent values of
the jet Doppler factor, $\delta \sim 25$, 
the magnetic field,  $B \sim 1$ Gauss, and the Lorentz factors 
of the electrons that radiate the bulk of the observed power, $\gamma_{\rm max}
\sim 10^3$.  The largest disparity arises
in the distance, $r$, of the dissipation region from
the central black hole (i.e., the location of the ``blazar emission zone''), 
and this points to differing origins of the seed photons that are 
Compton up-scattered to the $\gamma$-ray range:  UV disk photons
reprocessed in the broad line region vs. IR emission of the obscuring dust. 
In particular, Vercellone et al. claim $r \sim 3 \times 10^{16}$\,cm, while 
Sikora et al. argue for $r \sim 10^{19}$\,cm.  
As we discuss below, our detection of the spectral break
can put further constraints on the location of dissipation region,
but ultimately, time-resolved broad band spectroscopy will be crucial.

\newcommand{\E}{E}

\subsection{Implications of the $\gamma$-ray Variability on the Lorentz
  Factor of the Source of the $\gamma$-ray emission}

The $\gamma$-ray flux from 3C~454.3 is clearly variable
on time scales ranging from days to decades, with the average flux value 
measured during the LAT observations being somewhat higher than that found 
by AGILE and much higher than that measured by EGRET.  
Assuming a ``concordance'' 
cosmology ($\Omega_{\Lambda} = 0.73, \Omega_{\rm matter} = 0.27$, and 
H$_{\rm o} = 71$ 
km s$^{-1}$ Mpc$^{-1}$), we infer a luminosity distance of $d_L = 5.5$\,Gpc and 
an apparent isotropic monochromatic luminosity of
$L_{\E_0} \approx 4 \pi d_L^2\, (\Gamma_{\rm low} - 1) \,\E_0 \,
F_{\E > \E_0} \approx 2 \times 10^{48}$\,erg\,s$^{-1}$, where $\E_0 = 100$\,MeV 
and we have used the average 2008 August 3--September 4 flux of 
$F_{\E_0} = 3.0 \times 10^{-6}$\,ph\,cm$^{-2}$\,s$^{-1}$.
The bolometric luminosity is expected to be much higher than this
since the high-energy spectral component peaks below $100$\,MeV.
This finding, together with the observed rapid, large amplitude variability
measured in the $\gamma$-ray band, puts strong constraints on the
source parameters.  If a stationary source of a given luminosity in
$\gamma$-rays and X-rays were to be as compact as one would infer from
the variability data, the source would be optically thick to
e$^{+}$/e$^{-}$ pair production, assuming that the X-ray emission is 
co-spatial with the the $\gamma$--ray emission. This assumption is justified
by the simultaneity of the flaring activity in both 
bands as observed during our MW campaign, and in particular, 
as seen from our full Swift XRT soft X-ray TOO observations (2008 
July--September; Abdo et al. 2009, in preparation).   Similar simultaneous X-ray/$\gamma$-ray variability has been measured in the past 
for other sources, such as 3C~279 (Wehrle et al. 1998).  

One obvious solution to the problem of excess pair-production opacity 
is to invoke
relativistic motion and/or expansion of the source. Following the arguments 
given in Mattox et al. (1993, with corrections pointed out
in Madejski et al. 1996), and adopting the doubling time scale of 
$t_d \approx 2$\,days and the observed X-ray flux of $S_X \approx 
3 \times 10^{-11}$\,erg\,cm$^{-2}$\,s$^{-1}$ (as measured during our campaign) 
at the observed photon frequency $\nu_X \approx 10^{18}$\,Hz (corresponding to 
the photons that annihilate 
with the GeV $\gamma$-rays in the jet rest frame), 
we estimate the Doppler factor $\delta$ required 
for the photon-photon annihilation 
optical depth to be $\tau_{\gamma\gamma} \leq 1$.
With the derived relation $\tau_{\gamma \gamma} \sim \sigma_T \, d_L^2 \, S_X / 3 \, 
t_d \, c^2 \, \E_X \, \delta^4$, where we put the emission 
region linear size $R = c \, t_d \, \delta / (1+z)$, the source-frame
photon energy $\E'_X = (1+z) \, h \nu_X / \delta$, and the
intrinsic X-ray luminosity $L'_X = 4 \pi d_L^2 \, \delta^{-4} \, S_X$, 
one obtains $\delta \gtrsim 8$ (note here that the estimated Doppler factor
scales as $\delta \propto t_d^{-1/4}$).  Of course, omitting the requirement
of co-spatiality of the X-ray and $\gamma$-ray emission regions relaxes this 
limit.  Nonetheless, it is interesting to compare this constraint on the
Doppler factor with the estimate obtained from the VLBI superluminal motion, 
$\delta =  24.6 \pm 4.5$ (Jorstad et al. 2005). 
As long as the velocity of the VLBI jet is the same as the velocity of 
the outflow within the blazar emission zone, this implies 
that the photon-photon
annihilation effects involving the X-ray emission generated within
the jet are negligible.  

\subsection{Implications of the Complex $\gamma$-ray Spectrum}

The LAT data imply a more complex spectrum than a simple power-law,
indicating significant softening of the photon spectral index 
by $\Delta \Gamma\sim 1.2$ towards
higher energies.  This is not consistent with a spectral change of 
$\Delta \Gamma = 0.5$ that is 
expected from the typical ``cooling break'' associated with radiative losses.
The observed softening may instead be due to an intrinsic 
decline or break in the particle distribution.  It can also arise, at least 
partially, from ``environmental'' reasons, i.e., the underlying
photon spectrum is modified by the photon-photon 
pair production.   For 3C~454.3, this is 
unlikely to be due to the intergalactic diffuse background light 
(EBL) which would only affect the spectra from a source at $z \leq 1$ at energies 
$\geq 40$~GeV.  Also, as argued above, the pair-production effects internal
to the jet
are negligible, since the expected jet Doppler factor ($\delta \sim 25$) 
is much larger than that required for internal $\gamma$-ray 
transparency ($\delta \gtrsim 8$).  
A photon field that is external to the jet but
local to the blazar central engine would need to peak at $\sim 0.2$\, keV
in order to account for the observed spectral
softening at $\sim 2$\,GeV.   This requirement would exclude 
photons produced in the Broad Line Region (BLR) as the source of pair opacity
since that emission peaks in the UV.  However, the required X-ray photons 
could be produced 
in the innermost part of the putative accretion disk surrounding the 
central black hole and/or a hot corona above the disk.  

We do not know the shape of the intrinsic, unabsorbed
$\gamma$-ray spectrum, but the simplest assumption is that it is a
straight power-law, as is believed to be produced 
via 1st-order Fermi acceleration at shocks. 
Departures from such a spectral shape may involve both softening and 
hardening towards higher energies over the canonical power-law component, 
arising due to radiative particle energy losses and/or efficient 
stochastic acceleration acting near the shock front (see, e.g, 
Stawarz \& Petrosian 2008).  For a simple power-law,
we can estimate the opacity that the emerging 
$\gamma$-rays would encounter 
as a function of distance from the black hole, given the 
observed spectral break at $2$\,GeV (e.g., Reimer 2007).  
Assuming crudely that the break
corresponds to the location where the pair-production opacity 
$\tau$ is of order unity, we can estimate the {\em minimum} 
distance at which the $\gamma$-rays are produced.  
We consider the soft X-ray target photons at $\sim 0.2$\,keV 
as being produced either in the inner parts of an accretion disk 
(assumed here to follow the behavior of a Shakura-Sunyaev disk
spectrum) or in an X-ray emitting hot corona above the disk. For a black
hole mass estimate for 3C~454.3 of $4\times 10^9$ M$_{\odot}$ (Gu et al. 2001) 
and a mass
accretion rate of $\sim 10$\% the Eddington rate, the bolometric
(Shakura-Sunyaev) disk luminosity would be $L_{\rm bol}\sim
2\times 10^{46}$ erg s$^{-1}$.  This is in good agreement with reports
from Raiteri et al. (2007) of the quiet state UV-spectrum of 3C~454.3
that shows a rise in the SED, suggesting the onset of the ``big blue
bump'' thermal accretion disk spectrum. It is also in good agreement
with the luminosity estimate of the BLR from Pian
et al. (2006) of order $L_{BLR}\sim 3\times 10^{45}$ erg s$^{-1}$,
implying a typical covering factor of $\sim$ 10\%. The interaction of
jet GeV photons with soft X-ray photons originating in the inner parts
of an accretion disk implies photon-photon interactions at preferentially 
small angles. The angular dependence of the threshold target photon energy, 
$\E_{\rm thr}=(2m_ec^2)^2/(2E_\gamma(1-\cos{\theta}))$, 
where $\theta$ is the interaction angle, requires correspondingly higher 
densities of energetic 
target photons than are expected to be emitted by a standard 
Shakura-Sunyaev disk around a $4\times 10^9$ M$_{\odot}$ black hole 
for which the thermal emission is weak at 0.2 keV and above. 

On the other hand, quasars often show soft X-ray emission extending up 
to $\sim 100$ keV that is believed to arise from Comptonization by a hot coronal 
plasma above the disk. This mechanism creates a characteristic flat 
continuum that can typically be modeled as an $E^{-2}$ photon spectrum with an
exponential cut-off at $\sim 100$ keV.  Rapid variability of this flux indicates
that it is emitted within the regions closest to the black hole.  Adopting the 
typical values of the ratio $L_{\rm X} / L_{\rm bol}$ of 1--10\% 
(Laor et al. 1997), we find that
the condition corresponding to a unity optical depth for 2 GeV photons
occurs at $r \lesssim 10^{-3}$--$10^{-2}$\,pc, 
corresponding to $\sim 10$--$100$ gravitational radii for 
$M_{\rm BH} = 4 \times 10^{9}$ M$_{\odot}$.  Thus, if the observed 
spectral break at $2$\,GeV were due to photon-photon pair production 
on the accretion disc corona radiation, the implied location of the 
$\gamma$-ray emitting region should be extremely close to the black hole.  
Another scenario can be envisioned in which radiation from the accretion disk 
corona is subsequently Thomson-scattered towards the 
jet by the free electrons in the 
gas in the more distant broad line region.  However, this would 
require an unrealistically high product of covering fraction and 
Thomson optical depth that would not be consistent with the observed 
profiles of emission lines in AGN unless the electrons were located 
well within the broad-line region.  Regardless, neither direct nor BLR-scattered
accretion disk corona radiation are likely sources of the required pair 
opacity, because the reprocessed $\gamma$-radiation
should then escape as soft and strong X-ray emission produced in a
pair cascade developed in the intense photon field of the accretion 
disk/disk corona (Levinson \& Blandford 1995;  Ghisellini \& Madau 1996),
which is not consistent with the collected X-ray data in the
context of the broad-band spectrum (see. e.g., Fig. 3).

We thus suggest that the observed $2$\,GeV 
break in the $\gamma$-ray spectrum of 3C~454.3 is due
to an intrinsic break in the spectrum of radiating particles. 
In the context of models invoking Comptonization of external
photons, such a break
implies a characteristic Lorentz factor, $\gamma_{\rm el}$, for the
electrons radiating at the
break.  For inverse-Compton scattering of ``seed'' photons that have energy 
$\E_{\rm seed, rest}$ in the rest frame of the blazar, 
the observed $\gamma$-ray photon energy is given by
\begin{equation}
E_{\gamma} = E_{\rm seed, rest}\, \delta \, \Gamma_{\rm jet} 
\, (1+z)^{-1}\,\gamma_{\rm el}^2.
\end{equation}
Hence, the Lorentz factor of electrons radiating at $E_{\gamma} = 2$\,GeV is 
\begin{equation}
\gamma_{\rm el} 
\approx 6 \times 10^3 \, (\delta / 10)^{-1/2} \, (\Gamma_{\rm jet}/10)^{-1/2}
\, (E_{\rm seed, rest} / {\rm eV})^{-1/2}
\end{equation}
The likely sources of seed photons in blazars are 
either UV photons from the accretion disk that are reprocessed in the BLR 
and which comprise primarily Ly$\alpha$ photons with
$E_{\rm seed, rest} \sim 10$ eV, or IR emission from obscuring dust which have
$E_{\rm seed, rest}\sim 0.3$ eV. 
These seed photon energies yield
$\gamma_{\rm el} \sim 10^3$ and $\sim 6 \times 10^3$, 
for the cases of UV or IR seed photons, respectively, assuming the kinematic
jet parameters $\delta = 25$ and $\Gamma_{\rm jet} = 15$ as inferred from VLBI
observations.  In principle, for those values of $\gamma_{\rm el}$, the 
$\gamma$-ray radiation is close to the onset of the Klein-Nishina (KN) regime, 
but as has been argued by several authors (Zdziarski \& Krolik 1993; 
Moderski et al. 2005), such a KN cut-off might be less severe than expected 
in the simplest scenarios.  Regardless, we also comment here that such complex 
{\em intrinsic} $\gamma$-ray spectrum, if common, 
may present an obstacle towards the 
use of luminous blazars as ``white light'' sources to study the intergalactic 
diffuse UV/optical background radiation, first because it is not known whether the
break can be reliably modeled, but also because the steepening spectrum 
reduces the photon statistics at the highest $\gamma$-ray 
energies, where the opacity of the diffuse background light is most pronounced.  
Finally, it is interesting to speculate about the issue of contribution of 
$\gamma$-ray - luminous blazars to the extragalactic $\gamma$-ray background.  
Specifically, if luminous blazars make a significant contribution to this background 
and if a spectral break at a few GeV is a common feature in such blazars, 
then evidence of such breaks should be present in the extragalactic diffuse 
emission.  To settle this issue, sensitive, well-calibrated data extending 
to the highest energies accessible by Fermi are necessary, but detailed analysis of the
diffuse extragalactic background in Fermi data is yet to be performed.  

\section{Summary}

We report strong and variable $\gamma$-ray emission from the blazar
3C~454.3 measured by the LAT instrument onboard the Fermi Gamma-ray
Space Telescope.  The source, in a flaring/active state since 2000, 
was easily detected and showed rapid variability described as symmetric
flares with rise and fall time of $\sim 3.5$ days, reaching a peak flux
of $F_{\rm E>100 MeV}$ of about $1.2 \times 10^{-5}$ photons cm$^{-2}$
s$^{-1}$.  The time scales of those flares, coupled with the X-ray
luminosity of the jet, allow us to provide a lower limit on the
Doppler factor of the jet of $\delta > 8$, consistent with the values
inferred for much larger spatial scales with VLBI measurements.

We also find that the $\gamma$-ray spectrum of 3C~454.3 is not a
simple power-law, but instead, steepens towards higher energies.  A
good, but not unique, description of the spectrum is a broken power-law 
with photon indices of $\sim 2.3$ and $\sim 3.5$, 
below and above a break at $\sim 2$ GeV, respectively.  
This break might be due to
photon-photon absorption to pair production, but this would require
the region responsible for production of $\gamma$-ray flux to be 
sufficiently close to the accretion disk/black hole
system to produce spectral signatures of the reprocessed 
$\gamma$-rays in the X-ray photon energy range, which are not observed.  
Instead, we propose that an intrinsic break in the electron 
spectrum around electron energies $E_{\rm el}$ of $\sim 10^3 \, m_e c^2$ is 
a more likely explanation for the observed $\gamma$-ray spectral shape.  

\noindent {\bf Acknowledgments:}
The $Fermi$ LAT Collaboration acknowledges generous ongoing support 
from a number of agencies and institutes that have supported both 
the development and the operation of the LAT as well as scientific 
data analysis.  These include the National Aeronautics and Space 
Administration and the Department of Energy in the United States, 
the Commissariat \`a l'Energie Atomique and the Centre National 
de la Recherche Scientifique / Institut National de Physique 
Nucl\'eaire et de Physique des Particules in France, the 
Agenzia Spaziale Italiana and the Istituto Nazionale di 
Fisica Nucleare in Italy, the Ministry of Education, Culture, 
Sports, Science and Technology (MEXT), High Energy Accelerator 
Research Organization (KEK) and Japan Aerospace Exploration 
Agency (JAXA) in Japan, and the K.~A. Wallenberg Foundation, 
the Swedish Research Council and the Swedish National Space Board in Sweden.
Additional support for science analysis during the operations phase 
from the following agencies is also gratefully acknowledged: the 
Istituto Nazionale di Astrofisica in Italy and the K.~A. Wallenberg 
Foundation in Sweden for providing a grant in support of a Royal 
Swedish Academy of Sciences Research fellowship.  SLAC 
researchers acknowledge the support by the US Department of Energy 
contract DE-AC02-76SF00515 to the SLAC National Accelerator Laboratory.  
This research is partly based on observations with the 100-m telescope of the MPIfR
(Max-Planck-Institut f\"ur Radioastronomie) at Effelsberg. This paper is partly 
based on observations carried out at the 30-m telescope of IRAM, which is supported by 
INSU/CNRS (France), MPG (Germany) and IGN (Spain).  Finally, 
we also acknowledge many fruitful discussions with Drs. L. Stawarz and M. Sikora.

\clearpage

\begin{figure}
\epsscale{.80}
\plotone{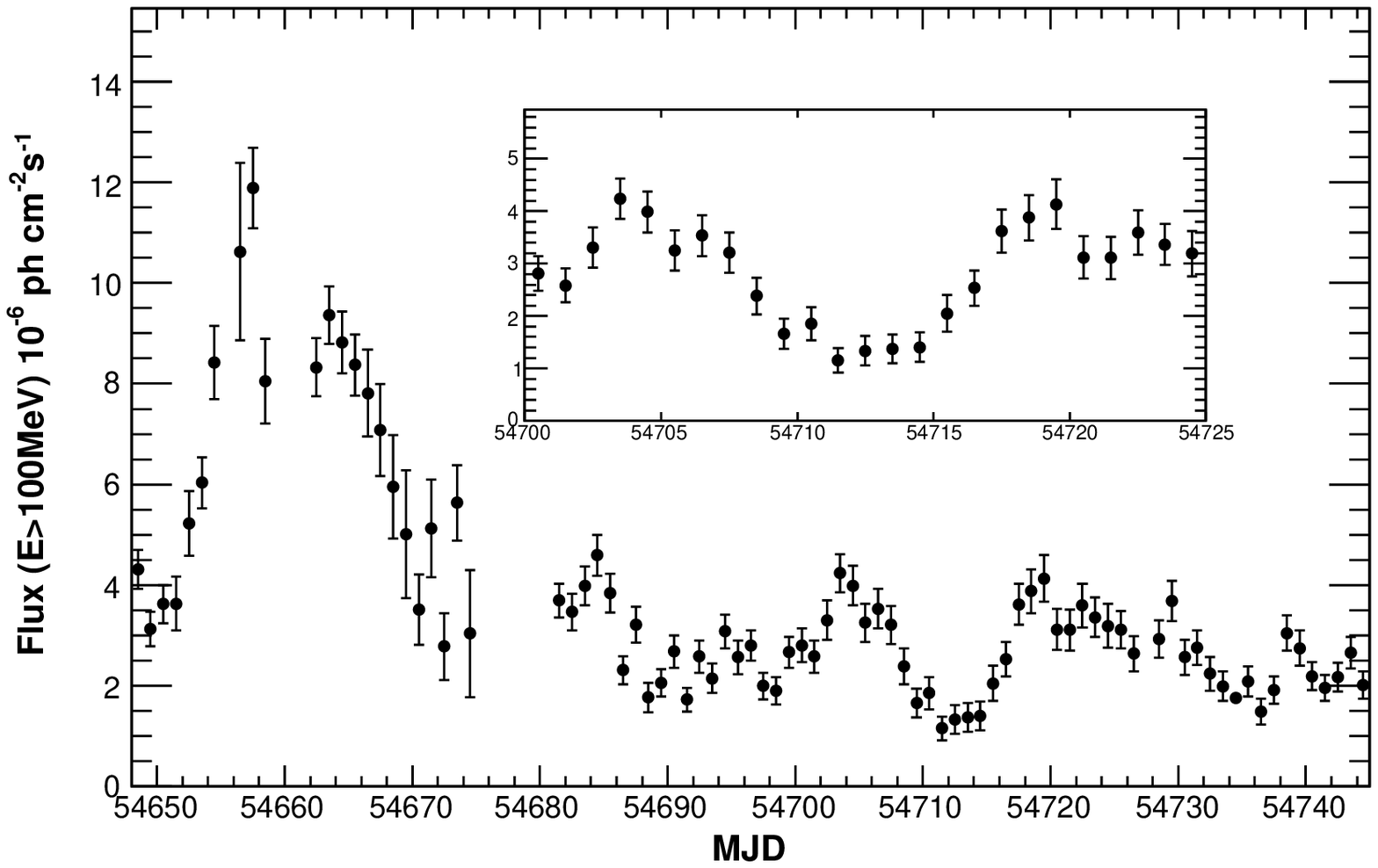}
\caption{ The flux light curve of 3C~454.3 in the
$100\,$MeV--$300\,$GeV band. The LAT operated in survey mode
throughout these observations except during the period MJD
54654--54681 (2008 July 7--August 2), when it operated in pointed
mode. The inset shows a blow up of the period MJD 54700-54725. 
The error bars are statistical only.
}
\label{fig:light_curve}
\end{figure}

\clearpage
\begin{figure}
\epsscale{.80}
\plotone{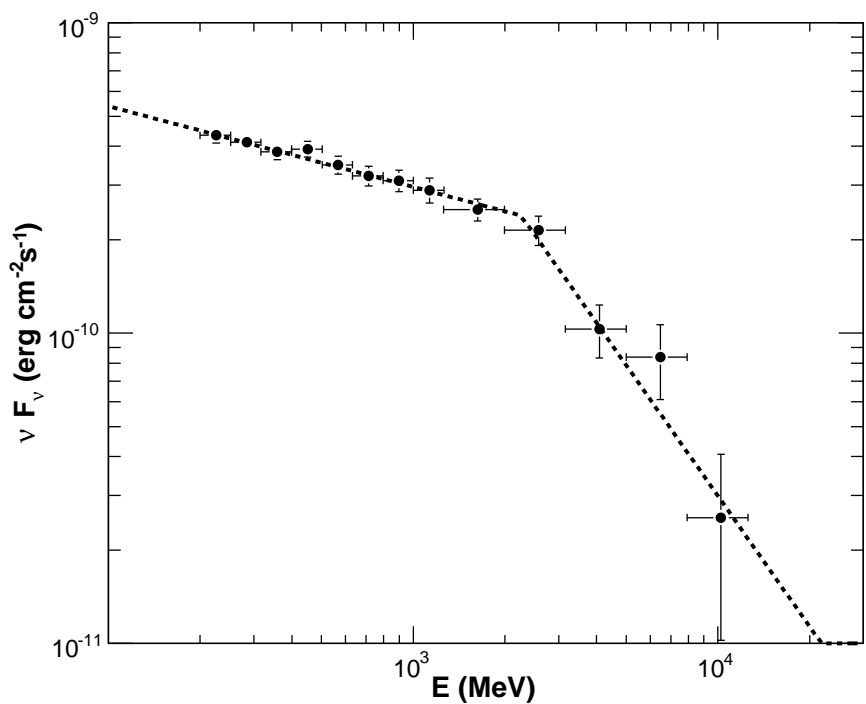}
\caption{  $\nu F_{\nu}$ distribution of the 
summed Fermi LAT data over the 2008 August 3--September~2 time 
span.  The model, fitted over the 200 MeV -- 300 GeV range, 
is a broken power-law with photon indices 
$\Gamma_{\rm low} = 2.27 \pm 0.03$, $\Gamma_{\rm high} = 3.5 \pm 0.3$, 
and a break energy $E_{\rm br} = 2.4 \pm 0.3$ GeV, and the apparent isotropic 
$E > 100$ MeV luminosity of $4.6 \times 10^{48}$ erg cm$^{-2}$ s$^{-1}$.  
The error bars are statistical only. 
\label{fig:LAT_nuFnu}}
\end{figure}

\clearpage
\begin{figure}
\epsscale{.80}
\includegraphics[angle=-90,width=15cm]{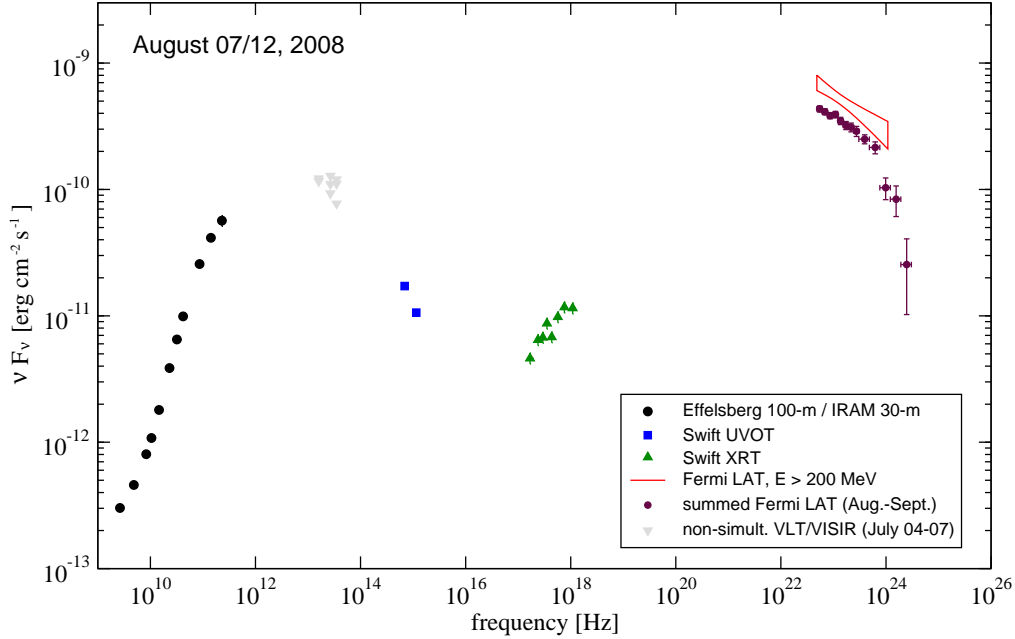}
\caption{Broad-band, (quasi-) simultaneous spectral energy
distribution of 3C~454.3 obtained during the period MJD 54685--54690 
(2008 August 7--12). The MW data are part of a larger follow-up 
campaign triggered by the early Fermi LAT results. Here, we show 
radio cm/mm band observations (2.6 to 230 GHz) obtained with the Effelsberg 
100-m and IRAM 30-m telescopes. The optical, UV and soft X-ray data were 
obtained from a 2 ksec pointing with the Swift satellite (UVOT, XRT) as part of 
daily TOO observations during the campaign (see Sect. 2.2 for details). 
For the $\gamma$-rays, the butterfly plot corresponds to the MJD 54685 
(2008 August 7) data, while the points correspond to the August data 
as in Fig. \ref{fig:LAT_nuFnu}. For illustration we also superimpose 
non-simultaneous mid-IR data obtained with the VLT/VISIR instruments 
during early 2008 July. Given the wide range of parameters that can 
reproduce the previously observed SEDs of 3C454.3 (see Section 3.1), 
we refrain from any detailed modeling, noting only that a hybrid 
synchrotron + external radiation Compton models can reproduce 
this SED adequately, and that the discrimination amongst details of 
various models requires time-resolved spectroscopy.  
\label{fig:SED}}
\end{figure}

\end{document}